\documentclass[mathleft
]{an}
\usepackage{graphicx}
\usepackage{times}
\usepackage{natbib}
\bibpunct{(}{)}{;}{a}{}{,}
\begin{document}

\Pagespan{789}{}
\Yearpublication{2006}%
\Yearsubmission{2005}%
\Month{11}%
\Volume{999}%
\Issue{88}%

\title{Local helioseismology and the active Sun}

\author{H. Schunker
\thanks{Corresponding author:
  \email{schunker@mps.mpg.de}\newline
  }
}
\titlerunning{Local helioseismology}
\authorrunning{H. Schunker}
\institute{
Max-Planck-Institut f\"ur Sonnensystemforschung, 
Max-Planck-Strasse 2,
Katlenburg-Lindau, 37191, Germany
}

\received{30 May 2005}
\accepted{11 Nov 2005}
\publonline{later}


\abstract{
The goal of local helioseismology is to elicit three-dimensional information about the sub-surface (or far-side) structure and dynamics of the Sun from observations of the helioseismic wave field at the surface. The physical quantities of interest include flows, sound-speed deviations and magnetic fields.  However, strong surface magnetic fields induce large perturbations to the waves making inversions difficult to interpret. The purpose of this paper is to outline the methods of analysis used in local helioseismology, review discoveries associated with the magnetic Sun made using local helioseismology from the past three years,  and highlight the efforts towards imaging the interior in the presence of strong magnetic fields.
}

\maketitle

\section{Introduction}
Helio- and astero-seismology provide the means to ``see" beneath the surface of the Sun or star, and to gain an understanding of the physics of stellar interiors. 
The analysis of the frequencies of the global modes of oscillation allows us to study the internal stellar structure only as a function of radius and unsigned latitude.
With high spatial resolution observations of the velocity signal on the Sun's surface, local helioseismology is able to image the three dimensional subsurface structure of the Sun and image magnetic activity on the far-side. 
Local helioseismology aims to interpret the full wave field observed at the solar surface, not only the frequencies of the normal modes.
We restrict our attention, here, to local helioseismology.

The Doppler signal, which is the line-of-sight velocity of the solar surface, is the fundamental observable. 
Full-disk high-quality data is obtained from instruments including the Michelson Doppler Image (MDI) onboard the Solar and Heliospheric Observatory (SOHO) \citep{MDI95}, the Global Oscillation Network Group  (GONG) \citep{Harveyetal1996}, and the Helioseismic and Magnetic Imager (HMI) onboard the recently launched Solar Dynamics Observatory (SDO). 
MDI also runs higher-resolution campaigns on smaller fields-of-view. 
The Magneto-Optical filter at Two Heights \citep{Finsterle2004} observes simultaneously at two heights in the atmosphere allowing an analysis of the vertical propagation of waves.

Various local helioseismic techniques exist to analyse the oscillations. 
The main ones are time-distance analysis \citep{DJHP93}, ring diagram analysis \citep{Hill1988} and acoustic imaging/holography \citep{CCL97,LB97}.  
Ring-diagram analysis is most closely related to global mode helioseismology. 
By analysing frequency shifts over small regions of the solar surface, the direction and amplitude of subsurface flows can be determined.
Time-distance analysis computes the travel time of a wave packet travelling between two points on the solar surface. Helioseismic holography (acoustic imaging) reconstructs the subsurface wave fields by propagating waves either forward or backward in time. 
Other methods include Fourier-Hankel analysis \citep{BDL87}, designed specifically to analyse waves surrounding a sunspot.
\cite{Woodard2009} used direct modelling to interpret the wave correlations in wavevector-frequency space.

The forward problem is then to compute the expected seismic observations from a particular model of the Sun. 
The inverse problem is to infer the internal properties of the Sun from the seismic observables.
Local helioseismology is the only way to elicit the three-dimensional structure of the solar interior, including structures and flows on a range of scales.

In this paper we focus on the local helioseismology of sunspots,  active regions and large scale flows varying with the solar cycle. 
There are particular problems associated with doing helioseismology in the immediate vicinity of strong surface magnetic fields.
The forward problem used to infer the subsurface properties has previously been based exclusively on solving hydrodynamic equations in the absence of a magnetic field. 
Additionally, the actual inversion method assumes that any perturbation to the waves is small, however it has been demonstrated that strong surface magnetic fields cause large perturbations to the waves.
Knowing the subsurface structure of sunspots and active regions will constrain sunspot models and hopefully determine their deeper structure.
The immediate need now is to quantitatively model wave propagation through magnetic fields and, subsequently, account for the effects when inverting for the subsurface structure. 

An important question in solar physics is to understand the mechanics of the solar dynamo.  Local helioseismology has identified large scale flow variations that appear to be intimately connected to the surface activity. 
Understanding such connections will help to decipher the larger scale dynamics of the magnetic Sun.
Other goals of local helioseismology are to measure the meridional flow and changes at the base of the convection zone, which would help constrain models of the dynamo.

This short review will cover the basic techniques used in local helioseismology  in Section~\ref{at}. 
How the interior structure is subsequently inferred is described in Section~\ref{inv}. Section~\ref{obs} will then cover some of the results, over the past three years, from local helioseismic analysis tied to solar activity. 
A concentrated effort towards modelling waves propagating through magnetic fields to  understand the observations will be reviewed in Section~\ref{mod}.
Recent reviews of local helioseismology include \cite{KHH06,GT07,Birch2008}.
For more extensive reviews of local helioseismology see, for example, \cite{GBS10}.

\section{Analysis techniques}\label{at}

\subsection{Ring-diagram analysis}
Typically ring-diagram analysis uses many small regions covering a significant fraction of the visible solar surface. 
Beneath each patch, two independent quantities, e.g. sound speed and density, may be inferred as a function of depth from which other quantities can be calculated. 
By combining the information in each patch, a three-dimensional picture is achieved. Typically this method is used to probe the top 30~Mm of the convection zone. 

For each patch a three-dimensional power spectrum is computed. 
A cut of the power spectrum at constant frequency results in an image of rings \citep{Hill1988}. 

Variations in the power spectra indicate particular subsurface properties. 
The presence of a flow shifts introduces a Doppler frequency shift and affects the shape of the rings. 
The degree to which the different modes are affected gives some indication of the depth and strength of the flow. 
Cuts at constant wavenumber can also be considered, where the power spectrum is a function of azimuthal angle and frequency. 
The measurement technique used in ring-diagram analysis and their interpretation are discussed in \cite[e.g.][]{BAT99,HHTBTH00}.

\subsection{Time-distance}
Time-distance helioseismology \citep{DJHP93} involves measuring the (phase) travel-time  between two points on the solar surface from the cross-covariance function of the Doppler signal at the two points.
The cross-covariance is sensitive to the wave speed in the solar interior.
For example, if there is a horizontal flow in the direction from point 1 to point 2  then this will result in a reduced travel time, and a longer travel time if the flow is in the opposite direction. 
One  way to extract the travel times from the cross-covariance is to fit Gabor wavelets to the cross-covariance \citep{Duvalletal1997} as a function of time-lag.
The cross-covariance can also be fit using a one parameter fit \cite{GB02,GB04}. 
Often, the cross-covariance is computed between a central point and a surrounding annulus, and averaged over the annulus. 
This technique is used to study the divergence of flows. 
The direction of the flows is gained by dividing the annulus into quadrants \citep[e.g.][]{DKSM96}.  

Filtering the observations is a common practice. In ridge filtering only modes of a particular radial order are retained, and in phase-speed filtering waves within a certain range of wave speed (or depth penetration) are retained.  
It has recently come to light that the filtering process has a strong effect on the travel-time measurements and must be included in any interpretation of the observations \citep{BB08,MHC09,Gizonetal2009}.

\subsection{Acoustic imaging/holography}

Acoustic imaging \citep{CCL97} and helioseismic \linebreak holography \citep{LB97} take the surface \linebreak Doppler observations and reconstruct the subsurface wave field from ingoing and outgoing waves with respect to a pupil. 
The ingoing waves are propagated forward in time and the outgoing waves are propagated backward in time using Green's functions and a knowledge of the solar interior.
Comparisons between the forward and backward propagating waves  are used to derive phase shifts and travel time variations. 
Far-side helioseismology  \citep{LB00} was developed using this method, although it is also possible with time-distance techniques \citep{Zhao2007}.

\subsection{Fourier-Hankel}\label{fh}
Specifically designed with the intention to study waves surrounding sunspots, Fourier-Hankel analysis \citep{BDL87} decomposes the observed oscillations into in- and out-going waves in  a cylindrical coordinate system centred on a sunspot. 
The observed waves are observed in an annulus surrounding the sunspot. 
From this the in- and out-going waves can be compared, in particular, their amplitude and phases \citep{BDL88,Braun1995}.

\section{Inversions}\label{inv}

The inverse problem in helioseismology involves extracting quantitative physical information regarding Solar subsurface inhomogeneities; such as sound-speed, density, mass flows, and magnetic field. 
This requires accurate and realistic forward modelling of the solar oscillations, given a prescribed
background model.
Model~S  \linebreak \citep{JCD96} is often used for this purpose in helioseismology. 
Furthermore, robust and reliable helioseismic travel-time measurements must be combined, or averaged, in some sensible way in order to best exploit the physical quantities of interest. 
It is also very important that the noise properties of the observations be well understood
and accounted for \citep{GB04}.

Heretofore, the effect of the inhomogeneities on the oscillations has been assumed to be weak. 
This ultimately yields a linear integral relationship relating observations, 3D sensitivity kernel functions, and the subsurface perturbations. 
As yet, no non-linear inversions have been attempted.

There exist two primary procedures for computing inversions for time-distance and ring-diagram analysis; The Regularised Least Squares (RLS) method \citep[e.g.][]{Kosovichev1996} and the Optimally Localised Averaging (OLA) \linebreak method \citep[e.g.][]{PT92,JGBT07}.

Accurate inversions depend heavily on the accuracy of the kernel functions, which must reflect all the aspects of the measurement procedure, including filtering. 
The ray approximation can be used to calculate the kernels \linebreak  \citep{Kosovichev1996}, but it is only useful if the background properties do not vary on scales smaller than, or comparable to, the wavelength. 
The first-order Born approximation offers a more reliable way to compute kernels, since it accounts for finite wavelength effects \citep{HDN01,GB02,BF04}.
Recently, \cite{BGHH07} have computed sensitivity kernels for ring-diagram analysis and \cite{BG07} have computed Born kernels for time-distance helioseismology. 

\cite{JGB08} have recently obtained maps of supergranular vector flows in the near-surface layers using an OLA inversion of travel times

\section{Recent observational results}\label{obs}

\subsection{Meridional flow}
Meridional flow is the flow from the Equator towards the poles near the surface, with a return flow at some depth. The meridional flow is important for magnetic flux transport in solar dynamo models.
An estimate of the meridional flow at the base of the convection zone predicts an amplitude of a few m s$^{-1}$ whereas the sound speed at this depth is a few thousand m s$^{-1}$, making it difficult to detect as discussed by \cite{BB08}.

Recently \cite{GR08} and \linebreak \cite{GKHHK08}  have measured the mean amplitude of the flow using time-distance and ring-diagram techniques respectively and get consistent values of $10-15$ m s$-1$ for the mean meridional flow near the surface.

\subsection{Solar cycle variations of meridional flow}
The meridional flow varies with the solar cycle. 
The residual from the mean flows at each latitude is a $\pm 5$ m s$^-1$ converging flow towards the active region belt  that increases in strength with increasing activity \citep{GR08,GKHHK08}. 

It is known that active regions are surrounded at the surface by a general inflow and when averaged longitudinally these flows introduce a perturbation to the meridional circulation.
\cite{Spruit2003} suggested that the inflow is due to the increased radiative loss of the active region compared to the surrounding Sun, and this can explain the solar cycle variations of the meridional flow.
In agreement with this, it has been shown by \cite{GR08} that a model including increased radiative loss in the active regions can reproduce the inflows and explain the observed surface solar-cycle variations.

By masking out active regions and their surrounding flows, \cite{Gizon2004} finds that the influence on longitudinal averages of active region flows is significantly reduced. This implies that the flows surrounding active regions are the major contributors to the time varying component of the meridional flow. 
Recently, \cite{GHHKH10} have shown that in this extended solar minimum the residual meridional flow developed before the onset of visible magnetic activity, suggesting that there is still a component independent of the presence of active regions.

\subsection{Torsional oscillations}
Torsional oscillations are bands of slightly faster rotation coinciding with the activity belt which migrate towards the equator over a solar cycle. The flow band associated with the new solar cycle has been found to be moving more slowly than in the lead up to the previous cycle \citep{Hetal2010}, possibly providing a clue as to why the onset of Solar Cycle 24 is later than expected.

\subsection{Sunspots}
There is considerable observational evidence that sunspots have a large effect on waves. It is well documented that sunspots absorb up to 50\% of the energy and shift the phase of incoming  waves \citep{BDL88,BLFJ92}.  Various dependencies on  frequency and radial order were also found. These results were originally discovered using \linebreak Fourier-Hankel methods. 

In the past few years, more evidence has emerged that the wave signal is dominated by surface effects \citep[e.g.][]{BB08,Gizonetal2009,Cally2009}.
Full waveform forward modelling has also revealed that a shallow sunspot model is sufficient to quantitatively explain the observed change in amplitude and phase of the waves in \linebreak caused by a sunspot \citep{CGSP10}. 
Despite the accumulating evidence that sunspots are large perturbations to the waves, wave-speed inversions below sunspots have traditionally been done under simplifying assumptions (weak and non-magnetic perturbations).
Such results should be \linebreak treated with caution due to the inherent problems of attempting  a  linear inversion involving non-linear perturbations to the waves.

Additionally, the inversions do not account for source suppression or details of the measurement procedure \cite[e.g.][show that the filtering is crucial]{BB08}.
Figure~19 from \cite{Moradietal2009} directly compares the results of various sunspot inversions.  
All of the methods, except the inversion for time-distance analysis, appear to be consistent with an increasing wave-speed in the top 2~Mm and small perturbations below this. 
However, this does not mean either result is correct and the cause(s) of the discongruity needs to be fully understood before a reliable inversion can be made.

For a more comprehensive review of sunspot seismology see e.g. \cite{Gizonetal2009}, and for sunspot models in local helioseismology \cite{Moradietal2009}.

\subsection{Acoustic power and magnetic activity}
It has been known for some time that acoustic power is altered by the presence of magnetic field. 
Suppression of acoustic power is evident in strong magnetic fields consistent with reduced convective sources. 
This suppression  affects the computed cross-covariances and thus the travel times.

It has also been observed that higher frequency \linebreak (5\,--\,6.5~mHz) wave power is enhanced in regions surrounding active regions and sunspots  \citep{Hindman:1998p398}, called acoustic halos. 
Recently, \cite{SB09} found that the halos tend to be stronger in regions of intermediate strength horizontal magnetic field.
From such a basic relationship between the acoustic power and magnetic field vector,  they are able to reproduce the acoustic power maps from the vector magnetic field to a high degree.
 They also found that the high frequency ridges of the power spectrum are shifted to higher wavenumber.
Several mechanisms have been suggested for generating the halo although none reproduce all of the observed characteristics of the halo. 
The mechanisms include mode conversion \citep{KC2009}, mode scattering \citep{Hanasoge2009}, trapping under magnetic field canopies \citep{Kuridze:2008p578}, and a change in convective scales \citep{JKWM08}.

 \cite{CYZLS09} analysed the spatial distribution of acoustic emission, absorption and suppression characteristics of two active regions and found that they all correlate spatially with the magnetic field, including the plage regions.

\cite{ZSET09} observe enhanced power in \linebreak sunspot umbra at large line-of-sight angles, i.e. perpendicular to the field. 
They model the propagation of waves \linebreak through localised magnetic field and conclude that it is \linebreak likely they are observing slow modes.

\section{Recent theoretical developments}\label{mod}
To understand the seismic signal observed on the surface of the Sun it is essential to model the propagation of waves through a magnetic field.
There are three ways that this has been done: 
\textit{i}) with simplified models aimed at understanding the physical mechanisms; 
\textit{ii}) with realistic numerical simulations of magnetoconvection where waves are naturally excited by convection; 
and \textit{iii}) with numerical simulations of linear wave propagation through a prescribed static background.

The models of \textit{i}) were instrumental for discerning the relevant physics. \cite{SB92} proposed that the absorption could be caused by mode conversion, and subsequent studies explored the physics \citep[e.g.][]{CB92,CB97}. 
In fact, \cite{CCCBD05} demonstrated that mode conversion could explain both the observed absorption and phase shifts and match the quantitative measurements quite well \citep{CCB03}. 
These models suggested that the mode conversion occurs in the top 1~Mm below the surface, where the Alfv\'en speed is comparable to the sound speed. More recently,  \cite{Cally2009} has shown that in umbrae thermal effects are likely to dominate the waves, whereas in the penumbra magnetic effect dominate.
 While these important results elucidate the mechanisms, now we require more realistic simulations where we can also test the analysis techniques and quantitatively measure the effects.

Approach \textit{ii}) is fully physical, but very computer intensive. 
These kinds of simulations allow a study of the behaviour of granulation, and thus changes in the properties of the waves, in the presence of magnetic fields. 
The most recent simulations including a sunspot can be seen in \cite{RSCK09}, although limited helioseismic analysis has been published \citep{BBR09}.

The third approach is less computer intensive and allows for parametric studies of magnetic structures.
A disadvantage is that these simulations require a stable background model. 
Several groups have developed numerical codes designed to model oscillations in the vicinity of sunspots using linearised MHD equations:
\begin{itemize}
\item The \textit{IAC MHD} code \citep{KCF08} has been used for a variety of studies. Some of the relevant results include that seismic waves below a sunspot travel faster than in the quiet-Sun and that high-frequency fast waves refracted by the magnetic field just above the surface are possibly the cause of acoustic halos \citep{KC2009}.
\item The \textit{SLiM} code  \citep{CGD07} together with a carefully constructed stable background \citep{SCG10} and sunspot model is able to reproduce observational results \citep{CGSP10}. The sunspot model itself was also somewhat constrained by the comparison \citep{CGD08}.  The code has also been used to study the separate effects of the magnetic field and thermal effects of a sunspot. 
\item The \textit{SPARC} code \citep{Hanasogeetal2006} has been used to study waves surrounding magnetic flux tubes, in particular the acoustic halo \citep{Hanasoge2009}. 
\item The \textit{SAC} code \citep{SFE08,SZFET09} was used to analyse the propagation and dispersion of waves in a non-uniform magnetic field. They also found that the largest contribution to travel time perturbations was due to the thermal structure in weak to intermediate magnetic field strengths, but that mode conversion in strong fields is important.
\end{itemize}
All of these codes give qualitatively similar results.

A careful analysis of the effects of stabilising a solar model against convection has been done using the SLiM code \citep{SCGM10}.  The stabilisation procedure changes the properties of the model, and in turn alters the eigenmodes.   \cite{SCGM10} attempt to correct for changes to the eigenmodes by modifying the background model.
 Since the models are stable, a model of the convective sources needs to be implemented as well as a model of the damping of the waves. 
 All of these parameters need to work in concert to produce seismically solar-like simulations to  provide realistic interpretations of the helioseismic observations.

\section{Discussion and future studies}

Further confirmation of the surface effects of strong magnetic fields have  been found in observations, casting further doubt on over-simplified inversions for wave-speed and flows below magnetic active regions and sunspots.
The effects of filtering, source suppression  and mode conversion of the waves in strong surface magnetic fields have yet to be fully quantified for each of the analysis techniques. 
In turn, a concerted effort is underway to quantitatively simulate the seismic signature of the Sun in the presence of magnetic fields to explore these effects. 
The simulations are still being fully analysed and in the near future a quantitative measure of the effects may be used to infer the correct subsurface properties of a sunspot.

Future efforts in local helioseismology will focus on deeper inversions, particularly of the base of the convection zone, and steps towards this  have been taken using time-distance techniques \citep{ZHKM09}.

A challenging goal is to detect the emergence of magnetic flux before it is observed at the surface. Some progress has been made towards this \citep[e.g.][]{KMHH08,KHH09,HKZM10}, but strong definitive signatures have not been observed.
Future studies of the effects of magnetic flux emergence will require systematic helioseismic analysis of many emerging active regions and realistic simulations.

The HMI instrument onboard the recently launched  \linebreak SDO  very soon provide increased resolution observations of the seismic Sun and provide accompanying vector magnetograms on a regular basis. Thanks to the improved spatial resolution we anticipate being able to do local helioseismology closer to the limb (and poles).


\acknowledgements
HS was supported by   the European Helio- and Asteroseismology Network (HELAS), a major international collaboration funded by the European Commission's Sixth Framework Programme.

\newpage
\bibliographystyle{aa}
\bibliography{papers_bib}

\begin{thebibliography}{72}
\expandafter\ifx\csname natexlab\endcsname\relax\def\natexlab#1{#1}\fi

\bibitem[{{Basu} {et~al.}(1999){Basu}, {Antia}, \& {Tripathy}}]{BAT99}
{Basu}, S., {Antia}, H.~M., \& {Tripathy}, S.~C. 1999, \apj, 512, 458

\bibitem[{{Birch} {et~al.}(2009){Birch}, {Braun}, \& {Rempel}}]{BBR09}
{Birch}, A., {Braun}, D.~C., \& {Rempel}, M. 2009, in AAS/Solar Physics
  Division Meeting, Vol.~40, AAS/Solar Physics Division Meeting

\bibitem[{{Birch}(2008)}]{Birch2008}
{Birch}, A.~C. 2008, Journal of Physics Conference Series, 118, 12009

\bibitem[{{Birch} \& {Felder}(2004)}]{BF04}
{Birch}, A.~C. \& {Felder}, G. 2004, \apj, 616, 1261

\bibitem[{{Birch} \& {Gizon}(2007)}]{BG07}
{Birch}, A.~C. \& {Gizon}, L. 2007, Astronomische Nachrichten, 328, 228

\bibitem[{{Birch} {et~al.}(2007){Birch}, {Gizon}, {Hindman}, \&
  {Haber}}]{BGHH07}
{Birch}, A.~C., {Gizon}, L., {Hindman}, B.~W., \& {Haber}, D.~A. 2007, \apj,
  662, 730

\bibitem[{{Braun}(1995)}]{Braun1995}
{Braun}, D.~C. 1995, \apj, 451, 859

\bibitem[{{Braun} \& {Birch}(2008)}]{BB08}
{Braun}, D.~C. \& {Birch}, A.~C. 2008, \solphys, 251, 267

\bibitem[{{Braun} {et~al.}(1987){Braun}, {Duvall}, \& {Labonte}}]{BDL87}
{Braun}, D.~C., {Duvall}, Jr., T.~L., \& {Labonte}, B.~J. 1987, \apjl, 319, L27

\bibitem[{{Braun} {et~al.}(1988){Braun}, {Duvall}, \& {Labonte}}]{BDL88}
{Braun}, D.~C., {Duvall}, Jr., T.~L., \& {Labonte}, B.~J. 1988, \apj, 335, 1015

\bibitem[{{Braun} {et~al.}(1992){Braun}, {Lindsey}, {Fan}, \&
  {Jefferies}}]{BLFJ92}
{Braun}, D.~C., {Lindsey}, C., {Fan}, Y., \& {Jefferies}, S.~M. 1992, \apj,
  392, 739

\bibitem[{{Cally}(2009)}]{Cally2009}
{Cally}, P.~S. 2009, \mnras, 395, 1309

\bibitem[{{Cally} \& {Bogdan}(1993)}]{CB92}
{Cally}, P.~S. \& {Bogdan}, T.~J. 1993, \apj, 402, 721

\bibitem[{{Cally} \& {Bogdan}(1997)}]{CB97}
{Cally}, P.~S. \& {Bogdan}, T.~J. 1997, \apjl, 486, L67

\bibitem[{{Cally} {et~al.}(2003){Cally}, {Crouch}, \& {Braun}}]{CCB03}
{Cally}, P.~S., {Crouch}, A.~D., \& {Braun}, D.~C. 2003, \mnras, 346, 381

\bibitem[{{Cameron} {et~al.}(2007){Cameron}, {Gizon}, \& {Daiffallah}}]{CGD07}
{Cameron}, R., {Gizon}, L., \& {Daiffallah}, K. 2007, Astronomische
  Nachrichten, 328, 313

\bibitem[{{Cameron} {et~al.}(2008){Cameron}, {Gizon}, \& {Duvall}}]{CGD08}
{Cameron}, R., {Gizon}, L., \& {Duvall}, Jr., T.~L. 2008, \solphys, 251, 291

\bibitem[{{Cameron} {et~al.}(2010){Cameron}, {Gizon}, {Schunker}, \&
  {Pietarila}}]{CGSP10}
{Cameron}, R., {Gizon}, L., {Schunker}, H., \& {Pietarila}, A. 2010, ArXiv
  e-prints:1003.0528

\bibitem[{{Chang} {et~al.}(1997){Chang}, {Chou}, {Labonte}, \& {The TON
  Team}}]{CCL97}
{Chang}, H., {Chou}, D., {Labonte}, B., \& {The TON Team}. 1997, \nat, 389, 825

\bibitem[{{Chou} {et~al.}(2009){Chou}, {Yang}, {Zhao}, {Liang}, \&
  {Sun}}]{CYZLS09}
{Chou}, D., {Yang}, M., {Zhao}, H., {Liang}, Z., \& {Sun}, M. 2009, \apj, 706,
  909

\bibitem[{{Christensen-Dalsgaard} {et~al.}(1996){Christensen-Dalsgaard},
  {Dappen}, {Ajukov}, {Anderson}, {Antia}, {Basu}, {Baturin}, {Berthomieu},
  {Chaboyer}, {Chitre}, {Cox}, {Demarque}, {Donatowicz}, {Dziembowski},
  {Gabriel}, {Gough}, {Guenther}, {Guzik}, {Harvey}, {Hill}, {Houdek},
  {Iglesias}, {Kosovichev}, {Leibacher}, {Morel}, {Proffitt}, {Provost},
  {Reiter}, {Rhodes}, {Rogers}, {Roxburgh}, {Thompson}, \& {Ulrich}}]{JCD96}
{Christensen-Dalsgaard}, J., {Dappen}, W., {Ajukov}, S.~V., {et~al.} 1996,
  Science, 272, 1286

\bibitem[{{Crouch} {et~al.}(2005){Crouch}, {Cally}, {Charbonneau}, {Braun}, \&
  {Desjardins}}]{CCCBD05}
{Crouch}, A.~D., {Cally}, P.~S., {Charbonneau}, P., {Braun}, D.~C., \&
  {Desjardins}, M. 2005, \mnras, 363, 1188

\bibitem[{{Duvall} {et~al.}(1993){Duvall}, {Jefferies}, {Harvey}, \&
  {Pomerantz}}]{DJHP93}
{Duvall}, Jr., T.~L., {Jefferies}, S.~M., {Harvey}, J.~W., \& {Pomerantz},
  M.~A. 1993, \nat, 362, 430

\bibitem[{{Duvall} {et~al.}(1997){Duvall}, {Kosovichev}, {Scherrer}, {Bogart},
  {Bush}, {de Forest}, {Hoeksema}, {Schou}, {Saba}, {Tarbell}, {Title},
  {Wolfson}, \& {Milford}}]{Duvalletal1997}
{Duvall}, Jr., T.~L., {Kosovichev}, A.~G., {Scherrer}, P.~H., {et~al.} 1997,
  \solphys, 170, 63

\bibitem[{{Duvall} {et~al.}(1996){Duvall}, {Kosovichev}, {Scherrer}, \&
  {Milford}}]{DKSM96}
{Duvall}, Jr., T.~L., {Kosovichev}, A.~G., {Scherrer}, P.~H., \& {Milford},
  P.~N. 1996, in Bulletin of the American Astronomical Society, Vol.~28,
  Bulletin of the American Astronomical Society, 898

\bibitem[{{Finsterle} {et~al.}(2004){Finsterle}, {Jefferies}, {Cacciani},
  {Rapex}, \& {McIntosh}}]{Finsterle2004}
{Finsterle}, W., {Jefferies}, S.~M., {Cacciani}, A., {Rapex}, P., \&
  {McIntosh}, S.~W. 2004, \apjl, 613, L185

\bibitem[{{Gizon}(2004)}]{Gizon2004}
{Gizon}, L. 2004, \solphys, 224, 217

\bibitem[{{Gizon} \& {Birch}(2002)}]{GB02}
{Gizon}, L. \& {Birch}, A.~C. 2002, \apj, 571, 966

\bibitem[{{Gizon} \& {Birch}(2004)}]{GB04}
{Gizon}, L. \& {Birch}, A.~C. 2004, \apj, 614, 472

\bibitem[{{Gizon} {et~al.}(2010){Gizon}, {Birch}, \& {Spruit}}]{GBS10}
{Gizon}, L., {Birch}, A.~C., \& {Spruit}, H.~C. 2010, ArXiv e-prints:1001.0930

\bibitem[{{Gizon} \& {Rempel}(2008)}]{GR08}
{Gizon}, L. \& {Rempel}, M. 2008, \solphys, 251, 241

\bibitem[{Gizon {et~al.}(2009)Gizon, Schunker, Baldner, Basu, Birch, Bogart,
  Braun, Cameron, Duvall, Hanasoge, Jackiewicz, Roth, Stahn, Thompson, \&
  Zharkov}]{Gizonetal2009}
Gizon, L., Schunker, H., Baldner, C.~S., {et~al.} 2009, Space Science Rev.,
  144, 249

\bibitem[{{Gizon} \& {Thompson}(2007)}]{GT07}
{Gizon}, L. \& {Thompson}, M.~J. 2007, Astronomische Nachrichten, 328, 204

\bibitem[{{Gonz{\'a}lez Hern{\'a}ndez} {et~al.}(2010){Gonz{\'a}lez
  Hern{\'a}ndez}, {Howe}, {Komm}, \& {Hill}}]{GHHKH10}
{Gonz{\'a}lez Hern{\'a}ndez}, I., {Howe}, R., {Komm}, R., \& {Hill}, F. 2010,
  \apjl, 713, L16

\bibitem[{{Gonz{\'a}lez Hern{\'a}ndez} {et~al.}(2008){Gonz{\'a}lez
  Hern{\'a}ndez}, {Kholikov}, {Hill}, {Howe}, \& {Komm}}]{GKHHK08}
{Gonz{\'a}lez Hern{\'a}ndez}, I., {Kholikov}, S., {Hill}, F., {Howe}, R., \&
  {Komm}, R. 2008, \solphys, 252, 235

\bibitem[{{Haber} {et~al.}(2000){Haber}, {Hindman}, {Toomre}, {Bogart},
  {Thompson}, \& {Hill}}]{HHTBTH00}
{Haber}, D.~A., {Hindman}, B.~W., {Toomre}, J., {et~al.} 2000, \solphys, 192,
  335

\bibitem[{{Hanasoge}(2009)}]{Hanasoge2009}
{Hanasoge}, S.~M. 2009, \aap, 503, 595

\bibitem[{{Hanasoge} {et~al.}(2006){Hanasoge}, {Larsen}, {Duvall}, {De Rosa},
  {Hurlburt}, {Schou}, {Roth}, {Christensen-Dalsgaard}, \&
  {Lele}}]{Hanasogeetal2006}
{Hanasoge}, S.~M., {Larsen}, R.~M., {Duvall}, Jr., T.~L., {et~al.} 2006, \apj,
  648, 1268

\bibitem[{{Hartlep} {et~al.}(2010){Hartlep}, {Kosovichev}, {Zhao}, \&
  {Mansour}}]{HKZM10}
{Hartlep}, T., {Kosovichev}, A.~G., {Zhao}, J., \& {Mansour}, N.~N. 2010, ArXiv
  e-prints:1003.4305

\bibitem[{{Harvey} {et~al.}(1996){Harvey}, {Hill}, {Hubbard}, {Kennedy},
  {Leibacher}, {Pintar}, {Gilman}, {Noyes}, {Title}, {Toomre}, {Ulrich},
  {Bhatnagar}, {Kennewell}, {Marquette}, {Patron}, {Saa}, \&
  {Yasukawa}}]{Harveyetal1996}
{Harvey}, J.~W., {Hill}, F., {Hubbard}, R.~P., {et~al.} 1996, Science, 272,
  1284

\bibitem[{{Hill}(1988)}]{Hill1988}
{Hill}, F. 1988, \apj, 333, 996

\bibitem[{Hindman \& Brown(1998)}]{Hindman:1998p398}
Hindman, B.~W. \& Brown, T.~M. 1998, Astrophys. J., 504, 1029

\bibitem[{{Howe} {et~al.}(2009){Howe}, {Christensen-Dalsgaard}, {Hill}, {Komm},
  {Schou}, \& {Thompson}}]{Hetal2010}
{Howe}, R., {Christensen-Dalsgaard}, J., {Hill}, F., {et~al.} 2009, \apjl, 701,
  L87

\bibitem[{{Hung} {et~al.}(2001){Hung}, {Dahlen}, \& {Nolet}}]{HDN01}
{Hung}, S., {Dahlen}, F.~A., \& {Nolet}, G. 2001, Geophysical Journal
  International, 146, 289

\bibitem[{{Jackiewicz} {et~al.}(2008){Jackiewicz}, {Gizon}, \& {Birch}}]{JGB08}
{Jackiewicz}, J., {Gizon}, L., \& {Birch}, A.~C. 2008, \solphys, 251, 381

\bibitem[{{Jackiewicz} {et~al.}(2007){Jackiewicz}, {Gizon}, {Birch}, \&
  {Thompson}}]{JGBT07}
{Jackiewicz}, J., {Gizon}, L., {Birch}, A.~C., \& {Thompson}, M.~J. 2007,
  Astronomische Nachrichten, 328, 234

\bibitem[{{Jacoutot} {et~al.}(2008){Jacoutot}, {Kosovichev}, {Wray}, \&
  {Mansour}}]{JKWM08}
{Jacoutot}, L., {Kosovichev}, A.~G., {Wray}, A., \& {Mansour}, N.~N. 2008,
  \apjl, 684, L51

\bibitem[{{Khomenko} \& {Collados}(2009)}]{KC2009}
{Khomenko}, E. \& {Collados}, M. 2009, \aap, 506, L5

\bibitem[{{Khomenko} {et~al.}(2008){Khomenko}, {Collados}, \& {Felipe}}]{KCF08}
{Khomenko}, E., {Collados}, M., \& {Felipe}, T. 2008, \solphys, 251, 589

\bibitem[{{Komm} {et~al.}(2006){Komm}, {Howe}, \& {Hill}}]{KHH06}
{Komm}, R., {Howe}, R., \& {Hill}, F. 2006, Advances in Space Research, 38, 845

\bibitem[{{Komm} {et~al.}(2009){Komm}, {Howe}, \& {Hill}}]{KHH09}
{Komm}, R., {Howe}, R., \& {Hill}, F. 2009, \solphys, 258, 13

\bibitem[{{Komm} {et~al.}(2008){Komm}, {Morita}, {Howe}, \& {Hill}}]{KMHH08}
{Komm}, R., {Morita}, S., {Howe}, R., \& {Hill}, F. 2008, \apj, 672, 1254

\bibitem[{{Kosovichev}(1996)}]{Kosovichev1996}
{Kosovichev}, A.~G. 1996, \apjl, 461, L55

\bibitem[{Kuridze {et~al.}(2008)Kuridze, Zaqarashvili, Shergelashvili, \&
  Poedts}]{Kuridze:2008p578}
Kuridze, D., Zaqarashvili, T.~V., Shergelashvili, B.~M., \& Poedts, S. 2008,
  Ann. Geophys., 26, 2983

\bibitem[{{Lindsey} \& {Braun}(1997)}]{LB97}
{Lindsey}, C. \& {Braun}, D.~C. 1997, \apj, 485, 895

\bibitem[{{Lindsey} \& {Braun}(2000)}]{LB00}
{Lindsey}, C. \& {Braun}, D.~C. 2000, Science, 287, 1799

\bibitem[{{Moradi} {et~al.}(2009{\natexlab{a}}){Moradi}, {Baldner}, {Birch},
  {Braun}, {Cameron}, {Duvall}, {Gizon}, {Haber}, {Hanasoge}, {Jackiewicz},
  {Khomenko}, {Komm}, {Rajaguru}, {Rempel}, {Roth}, {Schlichenmaier},
  {Schunker}, {Spruit}, {Strassmeier}, {Thompson}, \&
  {Zharkov}}]{Moradietal2009}
{Moradi}, H., {Baldner}, C., {Birch}, A.~C., {et~al.} 2009{\natexlab{a}}, ArXiv
  e-prints:0912.4982

\bibitem[{{Moradi} {et~al.}(2009{\natexlab{b}}){Moradi}, {Hanasoge}, \&
  {Cally}}]{MHC09}
{Moradi}, H., {Hanasoge}, S.~M., \& {Cally}, P.~S. 2009{\natexlab{b}}, \apjl,
  690, L72

\bibitem[{{Pijpers} \& {Thompson}(1992)}]{PT92}
{Pijpers}, F.~P. \& {Thompson}, M.~J. 1992, \aap, 262, L33

\bibitem[{{Rempel} {et~al.}(2009){Rempel}, {Sch{\"u}ssler}, {Cameron}, \&
  {Kn{\"o}lker}}]{RSCK09}
{Rempel}, M., {Sch{\"u}ssler}, M., {Cameron}, R.~H., \& {Kn{\"o}lker}, M. 2009,
  Science, 325, 171

\bibitem[{{Scherrer} {et~al.}(1995){Scherrer}, {Bogart}, {Bush}, {Hoeksema},
  {Kosovichev}, {Schou}, {Rosenberg}, {Springer}, {Tarbell}, {Title},
  {Wolfson}, {Zayer}, \& {MDI Engineering Team}}]{MDI95}
{Scherrer}, P.~H., {Bogart}, R.~S., {Bush}, R.~I., {et~al.} 1995, \solphys,
  162, 129

\bibitem[{{Schunker} \& {Braun}(2010)}]{SB09}
{Schunker}, H. \& {Braun}, D.~C. 2010, \solphys, accepted

\bibitem[{{Schunker} {et~al.}(2010{\natexlab{a}}){Schunker}, {Cameron}, \&
  {Gizon}}]{SCG10}
{Schunker}, H., {Cameron}, R., \& {Gizon}, L. 2010{\natexlab{a}}, ArXiv
  e-prints:1002.1969

\bibitem[{{Schunker} {et~al.}(2010{\natexlab{b}}){Schunker}, {Cameron},
  {Gizon}, \& {Moradi}}]{SCGM10}
{Schunker}, H., {Cameron}, R., {Gizon}, L., \& {Moradi}, H. 2010{\natexlab{b}},
  \solphys, submitted

\bibitem[{{Shelyag} {et~al.}(2008){Shelyag}, {Fedun}, \& {Erd{\'e}lyi}}]{SFE08}
{Shelyag}, S., {Fedun}, V., \& {Erd{\'e}lyi}, R. 2008, \aap, 486, 655

\bibitem[{{Shelyag} {et~al.}(2009){Shelyag}, {Zharkov}, {Fedun}, {Erd{\'e}lyi},
  \& {Thompson}}]{SZFET09}
{Shelyag}, S., {Zharkov}, S., {Fedun}, V., {Erd{\'e}lyi}, R., \& {Thompson},
  M.~J. 2009, \aap, 501, 735

\bibitem[{{Spruit}(2003)}]{Spruit2003}
{Spruit}, H.~C. 2003, \solphys, 213, 1

\bibitem[{{Spruit} \& {Bogdan}(1992)}]{SB92}
{Spruit}, H.~C. \& {Bogdan}, T.~J. 1992, \apjl, 391, L109

\bibitem[{{Woodard}(2009)}]{Woodard2009}
{Woodard}, M.~F. 2009, \apjl, 706, L62

\bibitem[{{Zhao}(2007)}]{Zhao2007}
{Zhao}, J. 2007, \apjl, 664, L139

\bibitem[{{Zhao} {et~al.}(2009){Zhao}, {Hartlep}, {Kosovichev}, \&
  {Mansour}}]{ZHKM09}
{Zhao}, J., {Hartlep}, T., {Kosovichev}, A.~G., \& {Mansour}, N.~N. 2009, \apj,
  702, 1150

\bibitem[{{Zharkov} {et~al.}(2009){Zharkov}, {Shelyag}, {Erd{\'e}lyi}, \&
  {Thompson}}]{ZSET09}
{Zharkov}, S., {Shelyag}, S., {Erd{\'e}lyi}, R., \& {Thompson}, M.~J. 2009,
  ArXiv e-prints

\end{thebibliography}

\end{document}